\documentclass[reprint,superscriptaddress,amsmath,amssymb,prapplied,aps,noeprint]{revtex4-2}

\usepackage{graphicx}
\usepackage{dcolumn}
\usepackage{bm}
\usepackage{braket}
\usepackage{hyperref}
\hypersetup{
    colorlinks=true,
    linkcolor=blue,     
    urlcolor=blue,
    citecolor=blue,
}
\usepackage[capitalise]{cleveref}
\usepackage{multirow}
\usepackage{tabularx}
\usepackage[euler]{textgreek}
\usepackage{xcolor}
\usepackage[version=4]{mhchem}
\usepackage{float}
\usepackage[mathlines]{lineno}
\newcommand{\NV}{N-\textit{V}}
\begin{document}
\title{Imaging the Meissner Effect and Flux Trapping of Superconductors under High Pressure using \NV Centers}

\author{Cassandra Dailledouze}
\affiliation{Université Paris-Saclay, CNRS, ENS Paris-Saclay, CentraleSupelec, LuMIn, F-91190 Gif-sur-Yvette, France 
}
\author{Antoine Hilberer}
\affiliation{CEA DAM, DIF, F-91297 Arpajon, France}
\affiliation{Université Paris-Saclay, CEA, Laboratoire Matière en Conditions Extr\^emes, 91680 Bruyères-le-Ch\^atel, France}
\author{Martin Schmidt}
\affiliation{Université Paris-Saclay, CNRS, ENS Paris-Saclay, CentraleSupelec, LuMIn, F-91190 Gif-sur-Yvette, France 
}
\author{Marie-Pierre Adam}
\affiliation{Université Paris-Saclay, CNRS, ENS Paris-Saclay, CentraleSupelec, LuMIn, F-91190 Gif-sur-Yvette, France 
}
\author{Loïc Toraille}
\affiliation{CEA DAM, DIF, F-91297 Arpajon, France}
\affiliation{Université Paris-Saclay, CEA, Laboratoire Matière en Conditions Extr\^emes, 91680 Bruyères-le-Ch\^atel, France}
\author{Kin On Ho}

\affiliation{Université Paris-Saclay, CNRS, ENS Paris-Saclay, CentraleSupelec, LuMIn, F-91190 Gif-sur-Yvette, France 
}
\author{Anne Forget}
\affiliation{SPEC, CEA, CNRS, Université Paris-Saclay, 91191 Gif sur Yvette, France
}
\author{Dorothée Colson}
\affiliation{SPEC, CEA, CNRS, Université Paris-Saclay, 91191 Gif sur Yvette, France
}
\author{Paul Loubeyre}
\affiliation{CEA DAM, DIF, F-91297 Arpajon, France}
\affiliation{Université Paris-Saclay, CEA, Laboratoire Matière en Conditions Extr\^emes, 91680 Bruyères-le-Ch\^atel, France}
\author{Jean-François Roch}
\email{jean-francois.roch@ens-paris-saclay.fr}
\affiliation{Université Paris-Saclay, CNRS, ENS Paris-Saclay, CentraleSupelec, LuMIn, F-91190 Gif-sur-Yvette, France 
}

\date{\today}

\begin{abstract}
Pressure is a key parameter for tuning or revealing superconductivity in materials and compounds. Many measurements of superconducting phase transition temperatures have been conducted using diamond anvil cells (DACs), which provide a wide pressure range and enable concomitant microscopic structural characterization of the sample. However, the inherently small sample volumes in DACs complicate the unambiguous detection of the Meissner effect, the hallmark of superconductivity. Recently, the Meissner effect in superconductors within a DAC was successfully demonstrated using diamond nitrogen-vacancy (\NV) widefield magnetometry, a non-invasive optical technique. In this work, we show that \NV magnetometry can also map superconductivity with micrometer resolution. We apply this technique to a microcrystal of \ce{HgBa2Ca2Cu3O_{8+\delta}} (Hg-1223) mercury-based cuprate superconductor under 4~GPa of pressure. The method is capable to detect the magnetic field expulsion and heterogeneities in the sample, visible in a set of characteristic parameters as the local critical temperature $T_{c}$. Flux pinning zones are identified through flux trapping maps. This approach could enable detailed investigations of superconductivity of a broad range of materials under high-pressure conditions.

\end{abstract}

\maketitle


\section{Introduction}

The Meissner effect, i.e. the expulsion of an external static magnetic field when a material is cooled below its critical temperature $T_{c}$~\cite{Meissner1933Neuer}, is the definitive signature of superconductivity. Understanding the behavior of this hallmark under extreme conditions, such as high pressure, is critical for exploring novel superconductors~\cite{Ashcroft1968Metallic,Pickard2020Superconducting,Lilia20222021,Sun2023Signatures}. Indeed, by directly modifying interatomic or intermolecular distances, pressure serves as a powerful tool for investigating new materials~\cite{Mao2018Solids}. High-pressure studies are also crucial in elucidating the mechanisms driving superconductivity, especially in unconventional systems~\cite{Mark2022Progress}. However, traditional methods, such as resistivity and magnetic measurements, are severely limited in their ability to study superconductivity under high pressure~\cite{Loveday2012High}. To date, performing reliable and reproducible magnetic measurements on samples compressed inside a diamond anvil cell (DAC) remains challenging due to the restricted sample access under pressure, the minute response from the micron-sized sample, and the poor spatial resolution and sensitivity of classical methods in these conditions. Moreover, in these measurements the magnetic signal from the sample is often polluted by the background coming from the pressure cell apparatus~\cite{Hirsch2022Faulty}. These challenges are further compounded by the extreme pressures required to stabilize some state-of-the-art superconductors, such as superhydrides above 100 GPa, where the sample can consist of multiple phases with varying stoichiometries and heterogeneities.

To tackle these challenges, recent studies proposed an optical magnetometry method based on diamond nitrogen vacancy (\NV) centers incorporated inside a DAC~\cite{Lesik2019Magnetic, Yip2019Measuring, Hsieh2019Imaging,Ho2021Recent,Bhattacharyya2024Imaging}. The \NV center, used as an \textit{in situ} quantum sensor, exhibits spin-dependent photoluminescence (PL) and is sensitive to an external magnetic field~\cite{Rondin2014Magnetometry}. At ambient pressure, \NV magnetometry can successfully capture several characteristic parameters of a superconducting state~\cite{Nusran2018Spatially, Joshi2019Measuring, Joshi2020Quantum}, and image vortices formation~\cite{Thiel2016Quantitative, Pelliccione2016Scanned, Schlussel2018Wide, Acosta2019Color, Lillie2020Laser, Nishimura2023Widefield,Hou2024probing}. Due to their intrinsic compatibility with the diamond anvil in a DAC, \NV centers offer an innovative approach to high-pressure experiments. Furthermore, it has been demonstrate that \NV optical magnetometry can be combined with synchrotron x-ray diffraction to simultaneously characterize the interlinked structural and magnetic properties of compounds in the DAC~\cite{Toraille2020Combined}. Additionally, the sensitivity of \NV centers to magnetic fields can be extended to pressures in the 100~GPa range ~\cite{Hilberer2023Enabling, Wang2024Imaging,Bhattacharyya2024Imaging}. The aim of the present study is to demonstrate the mapping with \NV centers of the Meissner expulsion in a bulk superconductor under high-pressure conditions.

Here we study, under $4$~GPa pressure, the Meissner effect of a \ce{HgBa2Ca2Cu3O_{8+\delta}} (Hg-1223) mercury-based cuprate sample using NV-based widefield magnetic imaging with micrometer resolution. As an alternative to the often difficult reconstruction of the full vector components of the distorted magnetic field~\cite{Chipaux2015Magnetic}, we introduce a signal analysis method to quantify parameters related to the splitting and spread of the NV electron spin spectrum. This approach allows for a qualitative visualization of the Meissner effect by focusing the analysis on any local modifications to the amplitude and direction of the measured magnetic field. This method provides fast and accurate analysis for large datasets. The mapping of the Meissner effect shows that the field expulsion extends over the entire sample, with its critical temperature $T_{c}$ value found to spatially vary by only approximately 4~K ($\Delta T_c /T_c < 3\%$). We also establish a flux trapping (FT) cartography, which highlights a key signature of superconductivity and indicates the presence of flux pinning centers caused by defects in the superconductor. We reveal a fine correlation between the magnetic and topographic properties of the sample using distinct cooling-down and warming-up protocols under pressure. Therefore, we show that widefield \NV magnetic imaging offers a robust determination of spatially resolved $T_{c}$ with inhomogeneities of the sample. These results pave the road to spatially resolved ultra-high pressure superconductivity research, without the need to reconstruct the full stray magnetic field created by the sample magnetization. This method will be particularly useful to understand mixed phases that can be formed in high-pressure novel material synthesis.

\section{Experimental setup and methods}

\begin{figure*}[t]
\includegraphics[width=\textwidth]{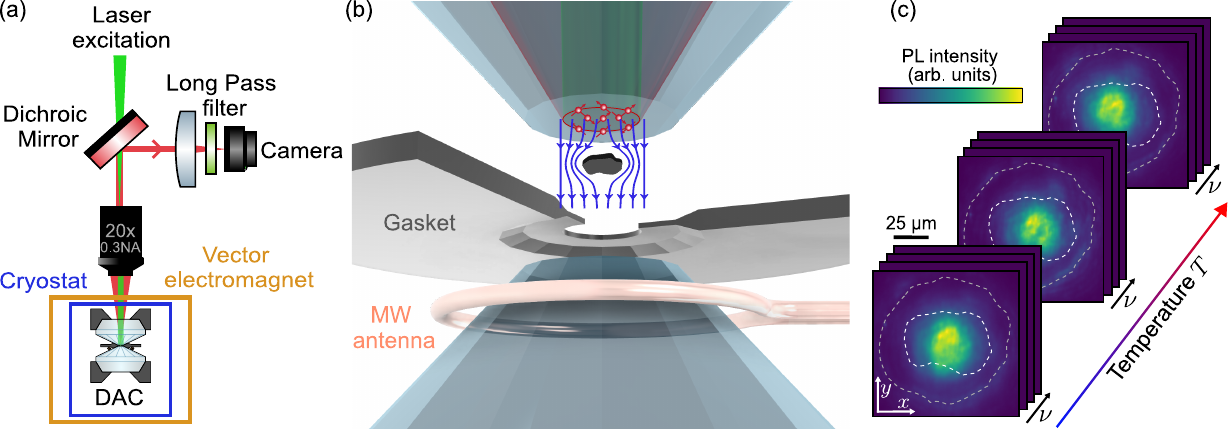}
\caption{(a) Experimental setup: Widefield \NV magnetic imaging inside a customized cryostat for spatially resolved detection of the Meissner effect. A $532$-nm wavelength laser serves as the excitation source, and the red PL emitted by the \NV centers is imaged through a long working distance microscope objective onto a CMOS camera. A vector electromagnet enables fine control of the external magnetic field used to induce the Meissner effect. (b) Illustration of the \NV-enabled DAC configuration for the Meissner effect detection. The superconductor expels the magnetic field lines when embedded in an external field, aligned along the DAC axis in our case. A single diamond anvil, implanted with \NV centers, allows the green laser excitation and the collected red PL to pass through the same side of the DAC. An external wire loop, coupled to a split gasket, provides MW excitation for the ODMR. (c) Schematic illustration of the widefield ODMR measurement protocol. The MW frequency $\nu$ is swept at a constant temperature and synchronized with the camera recording PL images, enabling the reconstruction of the ODMR spectrum for each pixel ($x,y$). This process is then repeated at different temperatures ($T$). The inside dashed line indicates the sample boundary, while the outside dashed line shows the gasket hole.}
\label{fig1}
\end{figure*}

The \NV center consists of a substitutional nitrogen atom at a carbon lattice site within diamond, paired with a vacancy at an adjacent crystallographic site. This system has spin-triplet ground state with sublevels $\ket{m_{s} = 0}$ and $\ket{m_{s} = \pm 1}$ that are zero-field split (ZFS) by $D=2.87$~GHz due to electron spin-spin interaction at ambient conditions, and without external perturbations. The \NV center can then reach corresponding spin excited states via green light absorption. The resulting spin-dependent red PL emitted by the \NV center can be used to measure the spin state by the so-called optically detected magnetic resonance (ODMR) technique~\cite{Gruber1997Scanning, Doherty2013Nitrogen}. Under continuous optical excitation, this quantum system polarizes into the $\ket{m_{s} = 0}$ state. Applying a microwave (MW) excitation matching the ZFS $D$ value can bring it into the $\ket{m_{s} = \pm 1}$ states. As the $\ket{m_{s} = \pm 1}$ states emit fewer photons, a decrease in the detected PL intensity indicates the spin transition. The frequency of the MW excitation is scanned across the resonance to obtain an ODMR spectrum which directly reveals the configuration of the triplet fine structure. Moreover, the degeneracy between the $\ket{m_{s} = + 1}$ and $\ket{m_{s} = - 1}$ states can be lifted by the Zeeman effect when an external magnetic field is applied~\cite{Rondin2014Magnetometry}. At first order, the observed frequency splitting between states $\ket{m_{s} = \pm 1}$ is proportional to the magnetic field projection along the \NV axis. 
In addition to the magnetic field, the effects of stress on the \NV center were shown to be significant in high-pressure experimental studies pioneered by Doherty \textit{et al.}~\cite{Doherty2014Electronic}. Strain environment models have been theoretically introduced to explain the observed stress effects on the electronic spin structure~\cite{Teissier2014Strain, Barson2017Nanomechanical, Broadway2019Microscopic, Barfuss2019Spin}: (1) the ODMR spectrum shifts to higher frequencies when a hydrostatic pressure is applied to the defect; (2) an additional splitting of the ODMR peaks arises under non-hydrostatic stress, which induces mixing between the $\ket{m_{s} = \pm 1}$ states. Therefore, it is crucial to carefully decouple these two external perturbations when doing high-pressure NV magnetometry.

The experimental setup is a widefield PL imaging microscope coupled to liquid $N_2$ cryostat, shown in \cref{fig1}(a) and described in more detail in the supplementary materials (SM) \cite{SM} and in previous studies~\cite{Lesik2019Magnetic,Hilberer2023Enabling}. The illustration of the \NV-enabled DAC is shown in \cref{fig1}(b). In particular, it takes advantage of the implantation of shallow ensembles of \NV centers into the single-crystal diamond anvil \cite{Lesik2013Maskless} which can enable the full vector magnetic field measurement through the simultaneous PL collection of the four families of NV centers \cite{Lai2009Influence}, distributed between the diamond crystal equivalent $[111]$ lattice directions. Our previous study~\cite{Hilberer2023Enabling} reaching up to $130$~GPa accounts for the influence of stress on the ODMR spectrum, thereby allowing the use of the \NV center zero-field ODMR spectrum as a direct pressure gauge. Argon is used as the pressure-transmitting medium \cite{Klotz2009Hydrostatic} and efficient MW excitation is achieved using the slit-gasket technique \cite{Meier2017Magnetic}. Finally, a thermocouple is placed close to the diamond anvil to provide precise temperature measurements.

Because of the Meissner effect, the superconductivity sample expels an applied magnetic field, as depicted in \cref{fig1}(b). The use of widefield microscopy to map this effect provides fast imaging of large areas with resolution close to the diffraction limit. The protocol to generate widefield \NV magnetic images is depicted in \cref{fig1}(c): For each scanned MW frequency, we record a camera image of the PL emitted by the \NV layer. Stacking all MW frequencies, we obtain a three-dimensional data volume containing PL images in one plane and ODMR spectra for each pixel in the transverse direction. This multiplexing in the acquisition time leads to a significant sensitivity advantage compared to, e.g., scanning optical confocal microscopy~\cite{Dreau2011Avoiding,Chipaux2015Magnetic,Scholten2021Widefield}. Finally, another dimension is added when the temperature is scanned to investigate superconductivity.

The mercury-bearing cuprate Hg-1223 is used as a testbed superconductor for our pressurized investigations, as excellent quality samples can be synthesized at ambient pressure with high reproducibility in the critical temperature~\cite{Loret2017Crystal}. Due to its high critical temperature $T_{c}$~\cite{Schilling1993Superconductivity, Nunez1993Pressure, Chu1993Superconductivity, Gao1994Superconductivity, Mark2022Progress}, the highest known before the recently discovered superhydrides, this three-layer cuprate allows experiments to be done using liquid nitrogen cooling. The cuprate sample used in this work is a microcrystal synthesized following the protocol described in~\cite{Loret2017Crystal}, and chosen with a surface area of approximately 50~\textmu $\text{m}^2$ and a thickness of about 15~\textmu m [\cref{fig2}(a)], fitting the dimensions of the DAC gasket hole. The bright field image of the cuprate embedded in the \NV-enabled DAC and the associated \NV widefield PL image are shown in \cref{fig2}(b).

We employ three well-known experimental protocols to study the Meissner effect of the Hg-1223 sample. (1) Zero-Field Cooling (ZFC-w): The sample is first ZFC to a base temperature of $90$~K, well below its estimated $T_c\sim140 $~K~\cite{Schilling1993Superconductivity, Nunez1993Pressure, Chu1993Superconductivity, Gao1994Superconductivity, Mark2022Progress}. Next, a magnetic field is applied and measurements are conducted during a warm-up phase. (2) Field cooling (FC-w): The sample is cooled to the base temperature with a magnetic field constantly applied, and measurements are conducted during a warm-up phase. (3) Field-cooled zero-field warming (or FT): The sample is cooled to the base temperature with a magnetic field constantly applied, then the magnetic field is removed and the remnant magnetic field created by the superconductor is measured during a warm-up phase. The ZFC-w and FC-w protocols are essential for studying the Meissner effect and determining the relevant superconducting critical parameters, while the FT protocol reveals another signature of superconductivity through the trapping of the initially applied external field by supercurrents around defects~\cite{Muller1987Flux}. 

\section{Analysis features of ODMR spectra}

\begin{figure}[t]
\includegraphics[width=8.6cm]{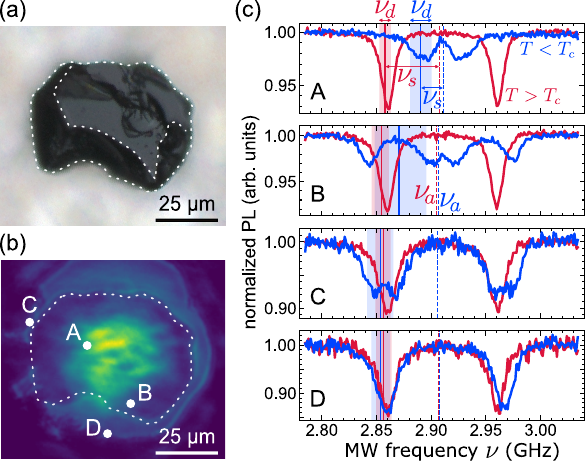}
\caption{(a) Bright field image of the Hg-1223 sample used in this work, taken outside of the DAC before loading under pressure. The dotted lines indicate both the outer boundary of the sample and the inner surface directly in contact with the diamond anvil. (b) Widefield PL image of the \NV centers in contact with the same sample under 4~GPa pressure in the DAC, obtained with green laser light excitation and recorded with an additional faint white light source. The shape of the sample can be seen in this image, allowing us to identify the position of the sample inside the gasket hole. (c) Selected ODMR spectra for different positions in the sample environment (A, B, C, and D) at two different temperature values under ZFC-w process. The red spectra are in the sample normal state taken at $T\sim150$~K, while the blue spectra are in the Meissner state taken at $T\sim100$~K. The position and temperature behaviors hint at a Meissner repulsion of the external magnetic field. The averaged frequency $\nu_a$ (dashed lines), the mean splitting $\nu_s$ (solid lines) and the spread $\nu_d$ (rectangles) are defined in the main text.
}
\label{fig2}
\end{figure}

\cref{fig2}(c) shows typical ODMR spectra below and above $T_c$ at four selected positions (A, B, C, and D) near the sample, obtained after a ZFC-w process. At both temperature, the same magnetic field $B = 3$~mT, aligned with the revolution axis of the cell ($[100]$ diamond direction), was applied. In the absence of magnetic expulsion from the cuprate above $T_c$, the four families of \NV centers feel the same projection of the magnetic field. In that case, the \NV response consists of two lines in the spectrum $\text{PL}(\nu)$ with summed high ODMR contrast, as illustrated in \cref{fig2} (c). When the superconductor begins to expel the magnetic field below $T_c$ as depicted in \cref{fig1} (b), this symmetry is broken. Depending on the position of the \NV centers above the superconductor, the orientation of the now perturbed magnetic field evolves differently. Therefore, a deviation of $\mathbf{B}$ from the $[100]$-axis lifts the degeneracy and separates the spin resonances of the four \NV families. This results in ODMR spectra where 2, 4, 6, and 8 peaks are visible in the ideal case. In \cref{fig2}(c), the ODMR spectra taken at temperatures $T<T_{c}$ illustrate this redistribution of the magnetic field due to the superconductor expulsion. Above the superconductor in position A, a strong reduction of the magnetic field is observed, evidenced by the lower Zeeman splitting of the ODMR lines. At the edges of the superconductor (positions B \& C), the $\mathbf{B}$ field is expected to rotate significantly from the [100] direction to accommodate field line redirections. This is indeed qualitatively observed in the increasing number of ODMR peaks (4 in the case of B \& C). Finally, at a certain distance from the superconductor, the magnetic field is unperturbed and matches that of the normal state. Note that these graphs directly demonstrate that the reconstruction of the full vector $\mathbf{B}$ field can be particularly difficult and inaccurate due to the numerous degeneracies hidden in the broadening of overlapping ODMR peaks.

Consequently, we employ a different method to further analyze the experimental data and outline the Meissner effect contribution to the ODMR signal. This method eliminates the need for classic fitting procedures of the ODMR spectra. To do so, we reduce the normalized PL spectra intensity $I(\nu)=1-\text{PL}(\nu)$ of each pixel and each temperature to a set of characteristic features, proposing a momentum analysis of the spectrum. 

First, we compute an averaged response frequency $\nu_a$ from the set of probed frequencies $\nu_i$ weighted by the recorded intensities $I(\nu_i)=I_i$ as
\begin{equation}
\label{nua}
    \nu_a=\frac{\sum_{i} \nu_{i}I_i}{\sum_{i} I_i}.
\end{equation}
This parameter directly measures the cumulative effect of the averaged local hydrostatic pressure around the \NV centers~\cite{Hilberer2023Enabling} and the off-axis magnetic field exerted on each \NV family~\cite{Tetienne2012Magnetic}. It corresponds practically to the middle spectral frequency usually defined in an ODMR spectrum where all the peaks are fitted, averaged over all families.

Next, we mirror the spectrum at $\nu_{a}$, leveraging its predominantly symmetric nature. In other words, we consider the frequency deviation $\tilde{\nu_i}$ from $\nu_{a}$ defined by $\displaystyle{\tilde{\nu_i}=\lvert \nu_i-\nu_a \rvert}$. From it, we calculate the mean frequency splitting $\nu_s$ of the PL intensity from the averaged frequency $\nu_{a}$ of the total spectrum
\begin{equation}
\label{nus}
    \nu_s=\frac{\sum_{i} \tilde{\nu_i}I_i}{\sum_{i} I_i}.
\end{equation}
This parameter $\nu_s$ measures the mean splitting of the resonance lines. It is thus related to the local magnetic field $\mathbf{B}$, which splits the resonant frequency of each \NV center family via the Zeeman effect depending on its relative orientation to $\mathbf{B}$. It corresponds roughly to half the splitting usually defined in an ODMR spectrum where all the peaks are fitted, but averaged over all families.

Finally, we estimate the variance of the half ODMR spectrum as the frequency spread $\nu_d$ of the resonances around the mean frequency splitting $\nu_s$ as
\begin{equation}
\label{nud}
    \nu_d=\frac{\sum_{i} \lvert \tilde{\nu_i}-\nu_s \rvert I_i}{\sum_{i} I_i}.
\end{equation}
A high spread $\nu_d$ means that the four \NV families experience significantly different projections of $\mathbf{B}$ onto their respective axes. The spread parameter $\nu_d$ is thus a measure of how much the local orientation of $\mathbf{B}$ differs from the initial DAC axis where $\mathbf{B}$ has the same projection on all \NV families. 

These three parameters are illustrated for different $\mathbf{B}$ field orientations in \cref{fig2}(c) with $T>T_{c}$ and $T<T_{c}$. The averaged frequency $\nu_a$ denoted with dashed lines is around $2.907 \pm 0.004 $~GHz in all cases. It is mainly determined by the local pressure $P$ which varies only slightly around 4~GPa. Near the center of the sample, corresponding to point A in \cref{fig2}(c), the evident reduction of the mean splitting $\nu_s$ at low temperature indicates a strong diamagnetism associated with the Meissner effect. Here, the degeneracies between \NV orientations start to lift due to the modification of the field orientation in the superconducting state, giving a larger spread $\nu_d$. Then, the magnetic field expulsion gradually weakens towards the border of the sample as observed at point B, and eventually slightly enhances the mean splitting $\nu_s$ near the exterior of the sample, as exemplified by point C. Close to the border [\cref{fig2}(c), point B \& C], the expulsion and rotation of the $\mathbf{B}$ field induces an increase of the spread $\nu_d$ compared to point A. At a finite distance outside the sample corresponding to ,e.g., point D, the diamagnetism associated with the Meissner effect is too weak to be observed, hence there is nearly no difference between the two spectra taken above and below  $T_{c}$. In that case, all four \NV axes are degenerate, resulting in a small spread $\nu_d$. 


To account for the significant variability in the magnetic field distribution without the need to individually distinguish between the various possible cases of $\nu_s$ and $\nu_d$ variations, we also introduce a general order parameter $\rho(T)$. This parameter provides a means for direct comparison between spectra and facilitates the determination of the critical temperature $T_c$ of the superconductor. $\rho(T)$ represents the correlation between a spectrum $I(T_{\text{ref}})$ at a reference temperature $T_{\text{ref}} \gg T_c$ and the spectrum $I(T,\nu)$ recorded at a given temperature $T$. The correlation is calculated using bivariate correlation in the form of Pearson's correlation coefficient~\cite{Pearson1896Vii,Dunn1986Applied,Lee1988Thirteen} (full expression
given in the SM~\cite{SM}). This parameter $\rho(T)$ is more general than the features defined above, as it captures any changes in the ODMR spectra that occur
when crossing $T_{c}$.

\section{Results}

\subsection{Micrometer-resolution spatial map of $T_c$}

Conventional high-pressure magnetic measurements methods provide only a bulk response of the sample. In contrast, the spatial resolution of the widefield \NV imaging method enables the mapping of the critical temperature $T_{c}$ across the entire sample.

\begin{figure}[t]
\includegraphics[width=8.5cm]{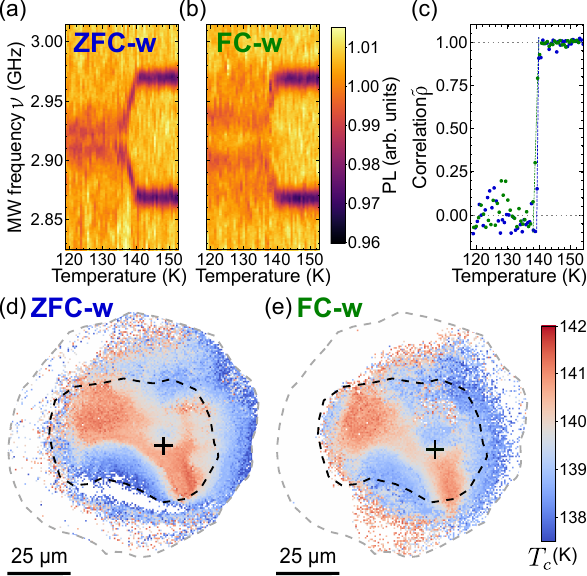}
\caption{(a, b) Colormaps of the \NV-ODMR signal using the ZFC-w and FC-w protocols, respectively, measured at the same point near the center of the superconductivity sample. These data were acquired at $4$~GPa under increasing temperatures with an external magnetic field of $3$~mT, for temperatures ranging from 120 to 150~K. Each vertical line represents an ODMR spectrum recorded at a specific temperature, with the PL intensity plotted in color. The transition from the superconductivity state to the normal state is clearly visible around 140~K. (c) Normalized correlation function $\tilde{\rho}$ as a function of temperature, derived from the data in panels (a) and (b). The ZFC-w and FC-w curves display a step-like transition, allowing the determination of $T_{c}$ as detailed in the text. Both protocols yield a $T_{c}$ value of $139.8 \pm 0.7$~K. (d, e) Extended $T_{c}$ maps obtained using the ZFC-w and FC-w protocols, respectively. The $T_{c}$ for each pixel has been determined using the normalized correlation parameter $\tilde{\rho}$. The dark dashed line indicates the sample boundary, while the gray dashed line outlines the gasket hole. The superconducting transition is observed in a region extending beyond the sample, as the Meissner expulsion effect extends to a finite distance outside the sample. The measurement location for (a) and (b) is marked by a cross.}
\label{fig3}
\end{figure}

Two series of ODMR spectra as a function of increasing temperature are shown in \cref{fig3}(a, b), representing PL data for the ZFC-w and FC-w protocols, respectively. These spectra were taken at a point near the center of the cuprate sample, with a low laser power to compromise between optimizing signal-to-noise ratio and minimizing parasitic heating~\cite{Lillie2020Laser}. The possible influence of laser and MW heating are discussed in the SM \cite{SM}. In these raw data, we can directly observe the superconducting transition by looking at the mean splitting $\nu_s$ or spread $\nu_d$ of the lines. These features here indicate that the magnetic field at the \NV centers position undergoes an abrupt transition from a Meissner-screened regime to a normal regime as the sample is heated. Interestingly, the behavior of the magnetic field is different between the ZFC-w and the FC-w protocol: At low temperature, the FC spectra seem to have a higher number of wider peaks compared to the ZFC-w spectra which seem to be comprised mainly of only two sharper peaks. These features seem to indicate that the magnetic field screening is superior in the ZFC-w case, and that the stray magnetic field is more rotated in the FC-w case. Nevertheless, the critical temperature seem identical, and can already be determined directly from the raw spectra. For adding precision, we use the correlation coefficient $\rho(T)$ previously defined. This calculation is particularly robust even for spectra resembling the FC-w spectra shown in \cref{fig3}(b) where the signal-to-noise ratio is significantly degraded at temperatures below $T_c$.

\cref{fig3}(c) shows the evolution of the correlation coefficient $\rho(T)$ with temperature for both acquisition protocols, calculated from the raw data shown in \cref{fig3}(a, b). The reference spectrum at high temperature used for the calculation, was obtained by averaging the ODMR spectra over a temperature range above $T_c$ from 145~K to 155~K. The correlation value was renormalized to $\tilde{\rho}$ using the median values at both ends of the whole temperature range studied here. A sharp step is observed in the correlation value at $T_c$, where the spectra abruptly change. To determine a precise value of the critical temperature $T_c$, we fit $\tilde{\rho}$ with a cumulative Gaussian function and identify the intersection between the tangent of this function at $\tilde{\rho}=0.5$ and the $\tilde{\rho}=1$ line, yielding $T_c=139.8 \pm 0.7$~K in this case. This $T_c$ value is expected, being in the range of temperatures determined by previous measurements at $P=4$~GPa with a bulk characterization~\cite{Schilling1993Superconductivity, Nunez1993Pressure, Chu1993Superconductivity, Gao1994Superconductivity, Mark2022Progress}. This correlation-based analysis provides a fast and efficient method to locate the sharp onset temperature of the superconducting transition, but is not aimed at analysing the transition temperature width.

This analysis is subsequently applied to all pixels of the images to convert the set of ODMR data into a full $T_{c}$ map [\cref{fig3}(d, e)]. Note that the determination of the critical temperature $T_{c}$ is based on the local distortion of the magnetic field due to the presence of the superconductor. Its influence extends beyond the strict geometric boundary of the sample, over almost the entire limit of the gasket hole. The obtained mapped area therefore exceeds the limits of the sample, shown as a dotted line in \cref{fig3}(d, e) and inferred from the optical image recorded in white light [\cref{fig2}(a, b)]. We designate those maps as \textit{extended} $T_{c}$ maps. 

Both ZFC-w and FC-w extended $T_{c}$ maps are similar. Each has a variation range of apparent $T_c$ for the cuprate crystal contained within $4$~K (between $137.5$~K and $141.5$~K), and regions in the sample where the superconducting state is maintained at a higher temperature during the warm-up phase. We confirmed that this difference of $\sim4$~K cannot be explained purely by heating artifacts (see the SM \cite{SM}). Therefore, the $T_{c}$ map indicates possible contributions from sample inhomogeneities.

To prove the presence of these inhomogeneities within the sample, it is necessary to go beyond the correlation measurement and examine the key features described previously to assess the magnetic field distribution. Notably, although $T_{c}$ appears to be the same for both ZFC-w and FC-w, the stray magnetic fields at low temperatures differ significantly between these two cases, motivating the following study.

\subsection{Magnetic field distribution and Meissner effect}

\begin{figure}[t]
\includegraphics[width=8.5cm]{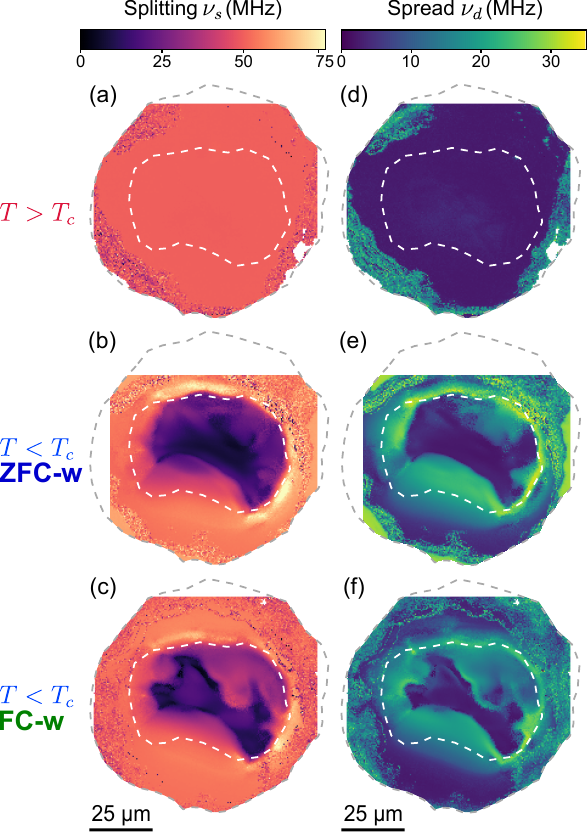}
\caption{(Left column a-c) Maps of the mean splitting parameter $\nu_s$ for (a) $T>T_{c}$, (b) $T<T_{c}$ using the ZFC-w protocol, and (c) $T<T_{c}$ using the FC-w protocol. The $T>T_{c}$ map is averaged over the temperature range $[142 \hspace{1mm}\text{K} ; 160 \hspace{1mm}\text{K}]$, while the $T<T_{c}$ maps are both averaged over $[90 \hspace{1mm}\text{K}; 110 \hspace{1mm}\text{K}]$. The white dashed line outlines the sample boundary, and the gray dashed line indicates the gasket hole. (Right column d-f) Maps of the spread parameter $\nu_d$ for (d) $T > T_c$, (e) $T < T_c$ using the ZFC-w protocol, and (f) $T < T_c$ using the FC-w protocol. The averaging temperature ranges for $\nu_d$ are identical to those used for $\nu_s$.}
\label{fig4}
\end{figure}

We now focus on analysing the stray magnetic field around the superconductor. The ZFC-w and FC-w protocols described earlier were repeated at the same pressure $P= 4$~GPa, and the same magnetic field $B = 3$~mT aligned with the DAC axis. Since the laser power has no significant influence on these measurements, a higher laser power was employed to maximize the \NV signal-to-noise ratio. From these measurements, we derive maps of the mean splitting parameter $\nu_s$ and spread parameter $\nu_d$ as defined in \cref{nus} and \cref{nud}, respectively. These maps are presented in \cref{fig4} and averaged over a temperature range of about 20~K in both cases, above and below $T_{c}$.

Above $T_c$, both $\nu_s$ and $\nu_d$ maps are uniform, serving as a reference to confirm the absence of the Meissner effect or any effect of the sample presence on the applied field.
Below $T_c$, the ZFC-w map reveals a significant decrease in the mean splitting $\nu_s$, indicating a reduction in the magnetic field near the sample due to strong magnetic field screening. The spatial evolution of the screening corresponds to the shape of the sample, as seen in \cref{fig2}(a), with a stronger screening observed in regions where the sample is closer to the \NV centers. A slight enhancement of $\nu_s$ is observed at the boundary of the sample, attributed to the concentration of magnetic flux in these regions where the expelled field lines have been redirected, as depicted in \cref{fig1}(b). In contrast, the FC-w map of $\nu_s$ shows a reduction in the Meissner expulsion in some areas in the sample, suggesting that some magnetic flux becomes trapped within the sample's defects. This behavior is characteristic of a type-II superconductor such as Hg-1223 and will be further discussed in the next section. 

The ZFC-w and FC-w maps of the spread parameter $\nu_d$ exhibit similarities to those of the mean splitting $\nu_s$ but highlight a different aspect of the flux expulsion process, as $\nu_d$ reflects the magnetic field orientation. Above the sample, where magnetic field screening is the strongest, the field orientation aligns with the applied field and thus the DAC axis, resulting in a minimal spread. At the edges of the sample surface nearest to the \NV centers, $\nu_d$ increases, signaling the gradual rotation of the magnetic field expelled by the sample. Consequently, the spread maps display sharper edges near the sample boundary compared to the $\nu_s$ maps, underscoring the utility of the spread parameter in visualising the spatial dependence of the Meissner effect.

To conclude, all these maps consistently reveal spatial inhomogeneities in the sample superconductivity behavior. Whereas the ZFC-w protocol predominantly indicates bulk expulsion of the magnetic field, the FC-w protocol highlights the existence of a penetrating magnetic field at localized, micrometer-scale  impurities. These impurities weaken the strength of the Meissner effect, as expected for a type-II superconductor. We now turn to examining how these impurities influence the FT behavior.

\subsection{Flux trapping}

\begin{figure}[t]
\includegraphics[width=8.5cm]{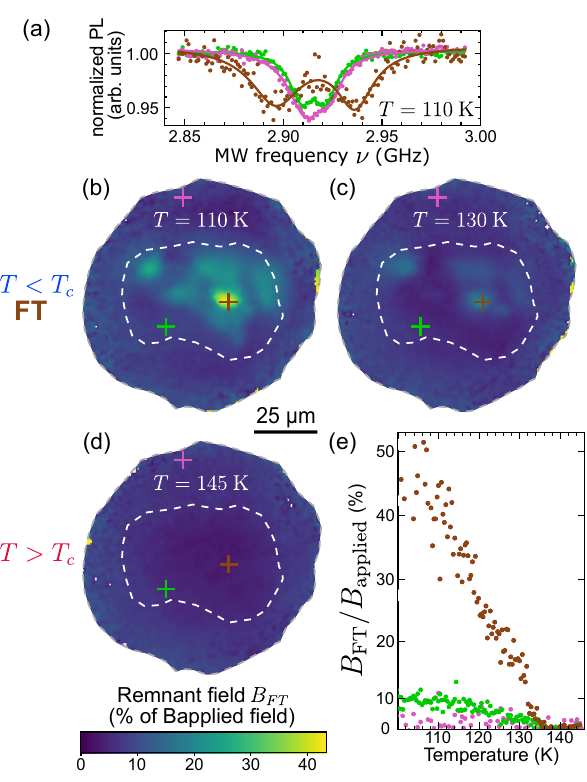}
\caption{(a) Selected ODMR spectra recorded in the FT protocol at different positions over the sample, indicated by crosses in panels (b)–(d) at $T=110$~K. The contrast of the brown ODMR spectra was multiplied by a factor $5$ for a better visualization. (b-d) Maps of the remnant magnetic field $B_{\rm{FT}}$ trapped within the sample, recorded at $T=110$~K, $T=130$~K, and $T=145$~K, respectively, expressed as a percentage of the initially applied magnetic field. The maps obtained for $T<T_{c}$ reveal a localized flux pinning zone. (e) Temperature evolution of the remnant field $B_{\rm{FT}}$ at three distinct locations, marked by crosses in panels (b)–(d).}
\label{fig5}
\end{figure}

In this section, we demonstrate the mapping of the FT effect. Spatially resolved studies of FT can reveal local defects in a given superconductor sample, and are interesting to investigate how these flux pinning centers evolve under pressure. As a proof-of-principle, we follow the FT protocol to obtain maps of the trapped magnetic flux at different temperatures, shown in \cref{fig5}. In the FT case, the trapped magnetic field is predominantly aligned along the DAC axis, resulting in only two peaks in the ODMR spectra as illustrated in \cref{fig5} (a). The orientation of the magnetic field remains relatively stable with temperature, allowing for accurate spectral fitting to extract a quantitative value for the remnant magnetic field $B_{\rm{FT}}$ defined as

\begin{equation}
    B_{\rm{FT}}=\frac{\sqrt{3}}{{2\gamma_{NV}}}\sqrt{\Delta^2-\Delta_\sigma^2},
    \label{eqFT}
\end{equation}
where $\gamma_{NV}= 28$~MHz/mT is the gyromagnetic factor of the \NV centers, $\Delta$ is the frequency splitting between the two ODMR peaks, and $\Delta_\sigma$ is the anisotropic stress splitting observed at $T>T_c$ when no trapped flux remains. This analysis accounts for the effects of pressure on the \NV centers as outlined in~\cite{Barson2017Nanomechanical,Broadway2019Microscopic}. Small deviations of the field orientation compared to the exact $[100]$-axis introduce minor variations in the apparent ODMR linewidth as shown in \cref{fig5}(a). This slight misalignment effect is further discussed in the SM \cite{SM}.

The maximal trapped magnetic field corresponds to approximately half of the initially applied magnetic field at the lowest measurement temperature of 110~K. The FT maps reveal that the localization of flux pinning centers correlates with the regions of flux penetration observed earlier in \cref{fig4}(c) and (f) under the FC-w protocol. Intuitively, the crack visible in the bright field image of \cref{fig2}(a) represents a significant extended defect in the sample, making it a natural flux pinning center. In \cref{fig5}(e), we plot the variation of the trapped flux as a function of temperature for various points near the sample. As expected, the trapped flux decreases as the temperature increases, reflecting the weakening of the superconducting screening currents responsible for the Meissner effect. These currents vanish in the normal state, leading to dissipation of the trapped flux upon warming.

\section{Conclusion and outlook}

In summary, we developed a widefield imaging technique to investigate superconductors under pressure with micrometer resolution, enabling spatially resolved studies of their behavior. Using this approach, we mapped the critical temperature $T_{c}$ via a correlation method and uncovered inhomogeneities within the superconducting sample. Additionally, we visualized the Meissner repulsion under pressure by introducing and mapping specially defined parameters: the mean frequency splitting $\nu_s$ and spread $\nu_d$, which efficiently quantify the evolution of the ODMR spectra. Furthermore, we compared the ZFC-w, FC-w, and FT protocols under pressure, constructing a cohesive picture of the sample’s topographic and magnetic properties. Similar \NV sensing protocols seems feasible under extreme conditions by carefully mitigating the non-hydrostatic strain exerted on the \NV centers~\cite{Hilberer2023Enabling, Ho2023Spectroscopic, Wang2024Imaging, Bhattacharyya2024Imaging}. \NV widefield imaging offers a promising platform for studying high-pressure superconductors with pronounced inhomogeneities, such as superhydrides synthesized under high-pressure-high-temperature conditions~\cite{Osmond2022Clean} and consisting of multiple phases with varying stoichiometries and heterogeneities.

By applying higher magnetic fields, lowering the temperature, and/or using a sample with a stronger demagnetization factor, it would be possible to generate spatial maps of the lower and upper critical fields, $H_{c1}$ and $H_{c2}$. Such maps could provide valuable insights into the spatial dependence of key superconducting parameters like the London penetration depth $\lambda$ and the coherence length $\xi$, as demonstrated in previous studies at ambient pressure~\cite{Joshi2019Measuring, Ho2024Studying} and high pressure~\cite{Yip2019Measuring}.

\section*{Acknowledgements}
We thank Florent Occelli for help in preparing and loading the low-temperature DACs, as well as Simon L'Horset and Sébastien Rousselot for technical support. We gratefully acknowledge Alain Sacuto and Jorge Hirsch for helpful discussions.

This work has received funding from the ANR with the ESR/EquipEx+ program (grant number ANR-21-ESRE-0031) and the SADAHPT program (grant number ANR-19-CE30-0027-01). JFR acknowledges support from the Institut Universitaire de France.

\bibliography{references}

\end{document}